\documentclass[runningheads]{llncs}
\usepackage{graphicx}
\usepackage{subcaption}
\usepackage[misc]{ifsym}
\usepackage{hyperref}
\usepackage{placeins}

\begin{document}
\title{Detection of vertebral fractures in CT using 3D Convolutional Neural Networks}
%
\titlerunning{3D CNN for vertebral fracture detection}
%

\author{Joeri Nicolaes\inst{1,2,3} \and
	Steven Raeymaeckers\inst{4} \and
	David Robben\inst{2,3,5}\and
	Guido Wilms\inst{6} \and
	Dirk Vandermeulen\inst{2,3} \and
	Cesar Libanati\inst{1} \and
	Marc Debois\inst{1}}

%
\authorrunning{J. Nicolaes et al.}
%
\institute{UCB Pharma, Brussels, Belgium \and
Medical Imaging Research Center (MIRC), KU Leuven, Leuven, Belgium \and
Medical Image Computing (MIC), ESAT-PSI, Department of Electrical Engineering, KU Leuven, Leuven, Belgium \and 
Department of Radiology, University Hospital, Brussels, Belgium \and
icometrix, Leuven, Belgium \and
Department of Radiology, UZ Leuven, Leuven, Belgium}

\maketitle              
\begin{abstract}
Osteoporosis induced fractures occur worldwide about every 3 seconds. Vertebral compression fractures are early signs of the disease and considered risk predictors for secondary osteoporotic fractures. We present a detection method to opportunistically screen spine-containing CT images for the presence of these vertebral fractures. Inspired by radiology practice, existing methods are based on 2D and 2.5D features but we present, to the best of our knowledge, the first method for detecting vertebral fractures in CT using automatically learned 3D feature maps. The presented method explicitly localizes these fractures allowing radiologists to interpret its results. We train a voxel-classification 3D Convolutional Neural Network (CNN) with a training database of 90 cases that has been semi-automatically generated using radiologist readings that are readily available in clinical practice.
Our 3D method produces an Area Under the Curve (AUC) of 95\% for patient-level fracture detection and an AUC of 93\% for vertebra-level fracture detection in a five-fold cross-validation experiment.

\end{abstract}
%
%
%

\setcounter{footnote}{0}

\section{Introduction}
\label{sec:intro}

Current radiology practice grades vertebral fractures according to Genant's semi-quantitative Vertebral Fracture Assessment (VFA) method~\cite{genant1993vertebral}. This method assesses the vertebral body morphology in X-ray images or at/around the mid-sagittal plane in 3D image modalities (CT, MR). As reported by Buckens et al.~\cite{buckens2013intra}, the intra- and inter-observer reliability and agreement of semi-quantitative VFA on chest CT is far from trivial on patient- and vertebra-level. 

A number of publications on vertebral fracture detection are inspired by how radiologists apply the Genant classification: firstly they attempt to segment the vertebrae at high accuracy, secondly the endplates are detected and finally the height loss of each vertebra is quantified in order to detect vertebral fractures. Such methods rely exclusively on 2D~\cite{Bromiley2016} and 2.5D~\cite{yao2012quantitative} height features.

Valentinitsch et al. propose a pipeline that first segments the vertebrae, then extract various 3D texture features (e.g. Histogram of Oriented Gradients,...) and volumetric Bone Mineral Density (vBMD) to finally apply a Random Forest classifier for patient-level fracture detection. Their experimental results show that combining multiple features calculated for each vertebra along the spine yields superior results~\cite{valentinitsch2019opportunistic}. Bar et al. does not first segment the spine before extracting features, but uses a Convolutional Neural Network (CNN) to directly map input images to output fracture classes. They combine a 2D CNN processing sagittal patches along the spine with a Recurrent Neural Network to aggregate predictions of multiple patches from the same patient. While this approach learns features from training data, it only uses 2D sagittal information at a virtually constructed sagittal section to cope with (abnormal) spine curvature~\cite{bar2017compression}. Tomita et al. apply a similar 2D approach using Long Short-Term Memory (LSTM) units for patient-level aggregation~\cite{tomita2018deep}.

In contrast, in this work we go beyond using learned 2D/2.5D or engineered 3D features by learning compact 3D features for detecting vertebral fractures. The proposed voxel classification method does not require segmenting the spine or (virtually) selecting the appropriate sagittal slice for inspecting vertebral fractures. 

\section{Data}
\label{sec:data}

\begin{figure*}
	\includegraphics[width=\textwidth]{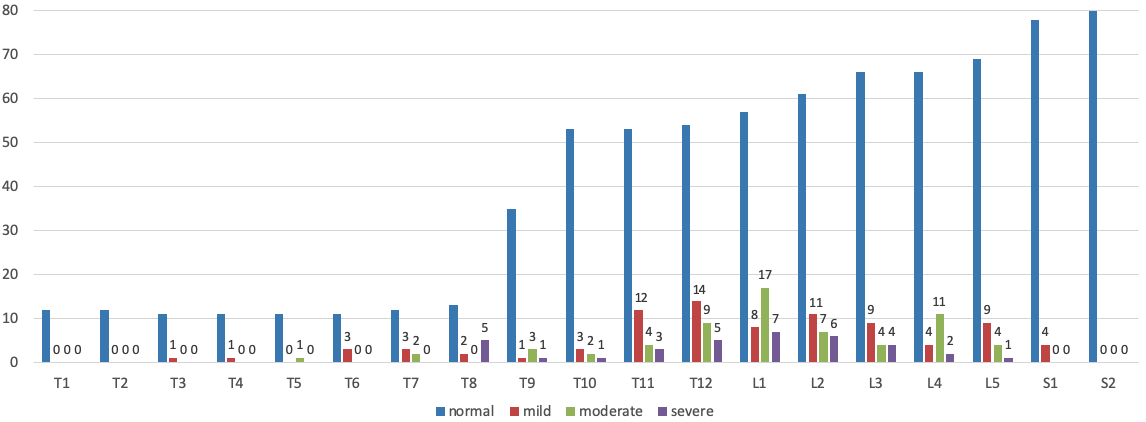}
	\caption{\protect\textbf{Training set}: number of fractures for every Genant grade along the spine, data labels indicate the amount of mild, moderate and severe fractures.}
	\label{fig:data}
\end{figure*}

For this study, we build a training database of 90 de-identified CT image series from the imaging database of the University Hospital of Brussels. These images were acquired on three different scanners (Siemens, Philips and General Electric; 120 kVp tube voltage; maximum in-plane spacing and slice thickness are respectively $0.92mm$ x $0.92mm$ and $1.5mm$) and contain 90 patients scanned for various indications (average age: 81 years, range: 70 - 101 years, 64\% female patients, 12\% negative cases). The dataset has been curated by one radiologist (S.R.) who scored Genant grades (normal, mild, moderate, severe) for every vertebra~\cite{genant1993vertebral}. It contains a total of 969 vertebrae of which 184 are fractured (85 mild, 64 moderate, 35 severe). Vertebral fracture prevalence is approximately 20\% in men and women above 60 years~\cite{waterloo2012prevalence} hence our dataset with 19\% vertebral fracture prevalence is representative for this clinical population. More than 90\% of the scans are abdomen studies implying that more than 75\% of the vertebrae range from T11 to S2\footnote{Vertebrae are named T1 to T12 for thoracic, L1 to L5 for lumbar and S1 - S2 for sacral vertebrae (with numbers increasing from top to bottom).}. Figure \ref{fig:data} shows the number of fractures for every Genant grade along the spine.

\section{Methods}
\label{sec:meth}

We present a two-staged vertebra fracture detection method that first predicts a class probability for every voxel using a 3D CNN and secondly aggregates this information to patient-level and vertebra-level fracture predictions. The CT images are resampled to $1 mm^3$ and normalized to zero mean and unit standard deviation before voxel classification.

\subsection{Voxel classification}
\label{sec:class}

Image classification CNNs map an input image to one output prediction for the entire image. This approach seems attractive for medical imaging as the expert labeling effort is limited to a (set of) answer(s) per image, yet building datasets of (tens of) thousands of CT scans containing subjects of the appropriate classes is not trivial. A voxel classification approach is typically applied for segmentation tasks. While this approach requires much less CT scans, it does require a label for every voxel in each training image which significantly increases the annotation effort. The proposed work applies a hybrid approach combining voxel classification with sparse vertebra annotations. Our experiments demonstrate that this approach produces good results using two orders of magnitude less images than a typical image classification approach. 

Since our task is detection and not segmentation, correctly predicting only a \textit{sufficient} amount of voxels around the vertebra centroid is needed to detect normal or fractured vertebrae in an image. We leverage this observation to construct 3D label images for our training database in a semi-automated fashion. First, radiologist S.R. created a text file with annotations for every vertebra present in the field of view as described in section \ref{sec:data}. Next, J.N. enriched these labels with 3D centroid coordinates by manually localizing every vertebra centroid in the image using MeVisLab~\cite{koenig2006mevislab}. This step required an average of less than two minutes per image in our dataset. Finally, we extended the method described by Glocker et al.~\cite{glocker2013vertebrae} to automatically generate 3D label images from these sparse annotations. 
The resulting label images contain ellipsoids (flattened along the longitudinal axis for fractured vertebrae) around each vertebra centroid annotated with the ground truth class label provided by the radiologist (combining mild, moderate and severe fractures into one fracture class because of the low number of examples per class, see Figure \ref{fig:data}). The generated label image is not \textit{voxel-perfect} under these assumptions as voxels near the vertebra border are labeled as background in the ground truth label image, but we demonstrate that this is sufficiently accurate for the fracture detection task at hand. 
The result of this step is a training database $\{ (I_k,L_k) \} _{k=1}^{K}$ with $K$ pairs of an image $I$ and label image $L$ of the same spatial dimensions that can be fed into a voxel-based CNN classifier. Note that the above semi-automated procedure is only required for building label images in our training database, test images are processed fully automatically by our method.

\subsection{CNN model selection}
\label{sec:model}

We know that human experts only leverage 2D height information from sagittal slices for detecting vertebral fractures (see section \ref{sec:intro}), but we want to investigate whether exploiting the 3D information in CT images does yield better results than only using 2D information in the sagittal plane. 

\subsubsection{Implementation details}
We used the open source Deepmedic Tensorflow implementation by Kamnitsas et al.~\cite{Kamnitsas201761} as this has proven to efficiently sample and process 3D segments from 3D images such as CT (since state-of-the-art GPUs cannot process full 3D image volumes due to memory constraints).

\begin{figure*}[!htb]
	\centering
	\includegraphics[width=\textwidth]{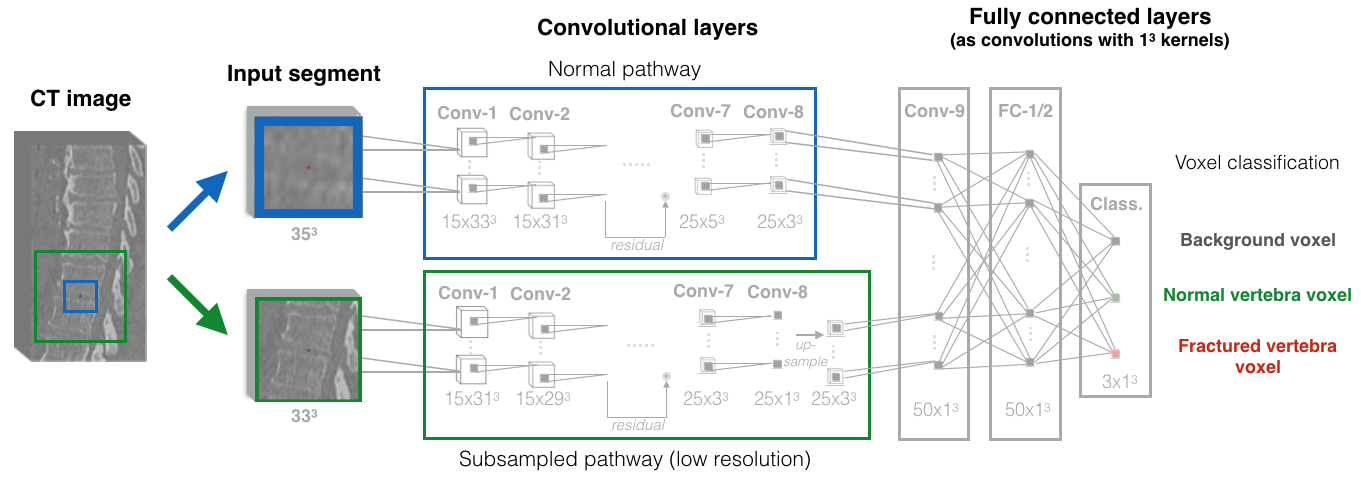}
	\caption{\protect\textbf{3D CNN} 11 layers dual pathway architecture realizes an effective receptive field of $17^3$ and $51^3$ in the normal and subsampled pathway respectively.}
	\label{fig:network}
\end{figure*}

All our experiments have been conducted using the voxel classification network shown in Figure \ref{fig:network}: an 11 layers dual pathway architecture containing 230K parameters. This CNN consists of 8 convolution layers each of which have filters of size $3^3$ realizing an effective receptive field of $17^3$ in the normal pathway and $51^3$ in the subsampled pathway (subsampling factor 3). This depth has been chosen such that features can be learned using all voxels inside a vertebral body. Additionally, we observed in our experiments that this effective receptive field yields distinct predictions for every vertebra. 

The following training regime has been used in all our experiments:
\begin{itemize}
	\item During training, image segments are sampled in a weighted regime using the ground truth label images to ensure that the network sees enough vertebra voxels. We apply a grid-sampling scheme during inference to build a prediction map of the entire image volume. 
	\item We apply data augmentation by adding noise to our input intensities and randomly flipping images across X, Y and Z axes. 
	\item We use the cross-entropy loss function, RmsProp optimizer, L1 and L2 regularization, anneal our initial learning rate of 0.001 when validation performance plateaus and train for 35 epochs.
\end{itemize}

\subsubsection{Model selection experiment} \label{sec:exp}
We investigated whether a 3D CNN performs better than a 2D equivalent by comparing three variants of the 11 layer dual pathway network. We split our training database randomly into training (N=68) and validation (N=22) to evaluate the three models depicted in Table \ref{tab:res}.
\begin{table}[h!]
	\caption{Model evaluation.}
	\label{tab:res}
	\centering
	\begin{tabular}[t]{ | c | c | c | c | }	
		\hline		
		\rule[-1.5ex]{0pt}{4.5ex} Model name & CONV-1 filter & CONV-2 to 8 filter & Receptive Field (normal pathway) \\
		\hline
		\rule[-1ex]{0pt}{3.5ex} \textbf{1slice} & 1x3x3  & 1x3x3 & $17^2$ in sagittal plane \\		\rule[-1ex]{0pt}{3.5ex} \textbf{5slices} & 5x3x3  & 1x3x3 & $17^2$ in sagittal plane \\
		\rule[-1ex]{0pt}{3.5ex} \textbf{3D} & 3x3x3  & 3x3x3 & $17^3$ in 3D \\		
		\hline
	\end{tabular}
\end{table}

\begin{figure*}[!htb]
	\centering
	\includegraphics[width=0.3\textwidth]{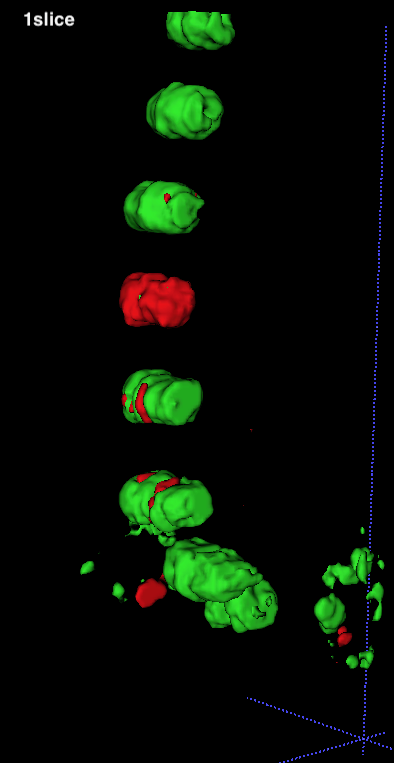}\hfill
	\includegraphics[width=0.3\textwidth]{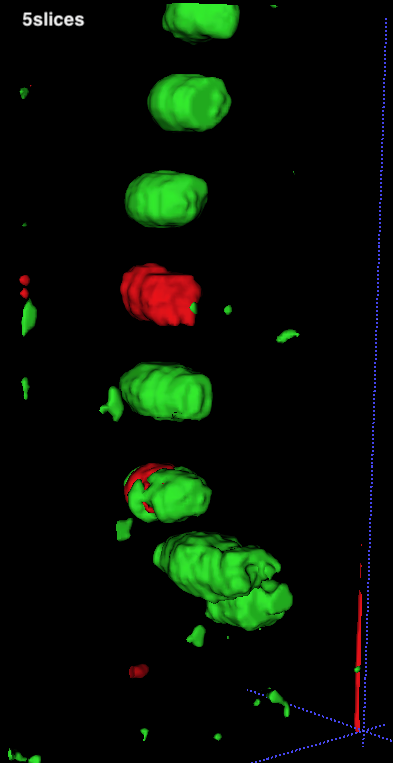}\hfill
	\includegraphics[width=0.3\textwidth]{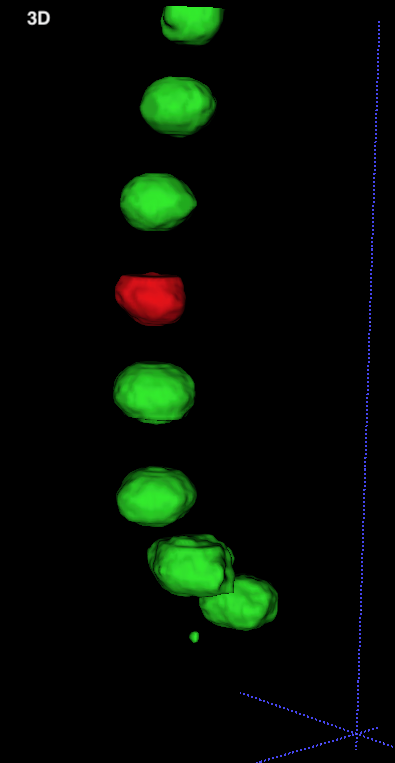}		
	\caption{\protect\textbf{Prediction outputs for 1slice, 5slices and 3D models} on one validation case in our training database. The network explicitly localizes the vertebra and labels each voxel with a class label (green = normal vertebra voxel, red = fractured vertebra voxel). The prediction images shown here are segmentation masks from epoch 30 using hard labels of the most likely class, rendered in 3D with a small counter-clockwise rotation around the longitudinal axis to show prediction results outside the sagittal plane. The 1slice (left) and 5slices (mid) models show stacked (sagittal) predictions. The 1slice (left) and 5slices (mid) models clearly yield more noisy predictions and mixed beliefs inside a vertebra while the 3D (right) model builds more compact predictions. Best viewed in color.}
	\label{fig:modelselection}
\end{figure*}

We calibrated all models to contain the same amount of network parameters and trained each model using the training regime described above. Every model has the same effective receptive field in the sagittal plane yet only the 3D model is allowed to learn features outside the sagittal plane. The 5slices model additionally learns to combine information from 5 input slices in the CONV-1 features.
Figure \ref{fig:modelselection} qualitatively compares the prediction results of these three models on one validation case. 
While all three models show similar prediction outputs at coarse scale, the 3D model clearly yields more compact and less noisy predictions than the 2D models at finer scale.
All subsequent experiments discussed in this work make use of the 3D model described in this section. 

\subsection{Aggregation}
\label{sec:aggr}
The voxel classifier transforms an input image into a prediction image that contains a probability $p(f|\mathbf{x})$, $f \in \mathcal{F} = \{\texttt{background}, \texttt{normal}, \texttt{fracture}\}$ for every voxel $\mathbf{x}$ present in the image. 
This information can be aggregated to patient-level (detecting whether a fracture is present in the patient image) or vertebra-level (detecting whether a fracture is present for every vertebra visible in the patient image).

\subsubsection{Patient-level fracture detection} \label{sec:patient}
First, we aggregate the 3D prediction image to patient-level fracture predictions by finding the connected components of fracture voxels and counting the total number of fracture voxels present in the image. 
This coarse form of aggregation involves two negatively correlated hyperparameters: a \textit{probability threshold} for selecting only those fracture voxels that have been predicted with high probability by our voxel classifier and a \textit{noise threshold} for determining when a component is too small to be a group of vertebra voxels. 

\subsubsection{Vertebra-level fracture detection} \label{sec:vert}
Secondly, we used the ground truth centroid coordinates that were annotated for building our training database (see section \ref{sec:class}) to perform a more fine-grained aggregation at vertebra-level. The \texttt{fracture} prediction probabilities of voxels inside a cube around the vertebra centroid are averaged to produce one summary score per vertebra. These probabilities are weighted using a Gaussian distance kernel to decrease the contribution of the voxels further away from the centroid (consistent with our ground truth label images which are \textit{by design} less accurate for voxels that are more distant from the ground truth centroid).
For automated screening, we envisage combining our vertebra-level fracture detection method with a vertebra localization method that automatically identifies and localizes each vertebra present in the image. Current state-of-the-art vertebra localization work reports identifying 91.6\% of the vertebrae and localizing them with mean error 6.2$\pm$16.2 mm~\cite{mader2017detection} on a challenging public dataset. We have used these localization error bounds to add noise to our ground truth centroid coordinates for simulating automated results.

\section{Results}
\label{sec:eval}


We performed a stratified 5-fold cross-validation\footnote{Since our training database has only 11 negative cases, we stratified the random sampling to ensure that each fold has a minimum of two negative cases.} using 90 images in our training database to estimate the expected performance of our 3D method.  
For each run, we selected 15\% of the images in the training folds as validation samples to determine when to stop training based.  We report the Receiver Operating Characteristic (ROC) curve because this metric describes model performance independently of the class distribution and is best suited to compare results from different test sets. The vertebra-level hyperparameter \textit{cube size} has been determined using cross-validation (10 voxels).

\begin{figure*}[!htb]
	\centering
	\includegraphics[width=\textwidth]{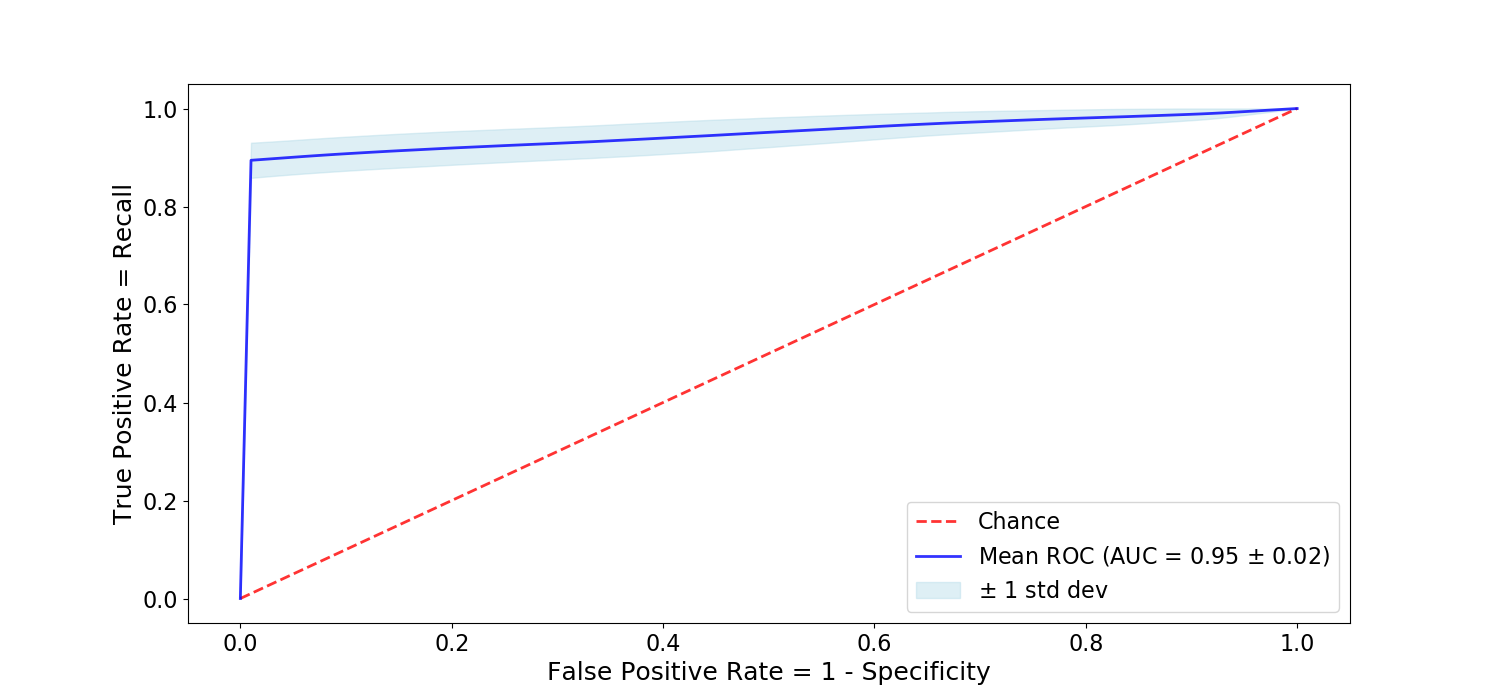}
	\caption{\protect\textbf{Patient-level fracture detection ROC curve} for the 5-fold cross-validation experiment: bootstrapped (n=1000) ROC curve (AUC=$0.95\pm0.02$) of our method in blue. }
	\label{fig:roc_p}
\end{figure*}

Since our \textbf{patient-level} fracture detection method involves two hyperparameters that can be chosen to deliver distinct classifiers, we build the ROC curve using the convex hull representing the optimal classifiers from a group of potential classifiers~\cite{provost1997analysis}. Each point on this ROC curve represents one optimal classifier generated with one pair of hyperparameter values (probability threshold, noise threshold). Figure \ref{fig:roc_p} shows this patient-level fracture detection ROC curve for the five-fold cross-validation experiment\footnote{The (False Positive Rate, True Positive Rate) values have been interpolated to plot a smoother curve.}. 
Our patient-level fracture detection Area Under the Curve (AUC) of $0.95\pm0.02$ is comparable to the results reported by Valentinitsch et al. (AUC 0.88)~\cite{valentinitsch2019opportunistic}, Tomita et al. (AUC 0.92)~\cite{tomita2018deep} and the operating point (recall 0.905, specificity 0.938) on our patient-level fracture detection ROC is similar to the one reported by Bar et al. (recall 0.839, specificity 0.938)~\cite{bar2017compression}. We note that all these results have been reported using different test sets (due to the absence of a public test set for fracture detection). We did not evaluate these other methods on our test set due to the absence of an open source implementation. 

\begin{figure*}
	\includegraphics[width=\textwidth]{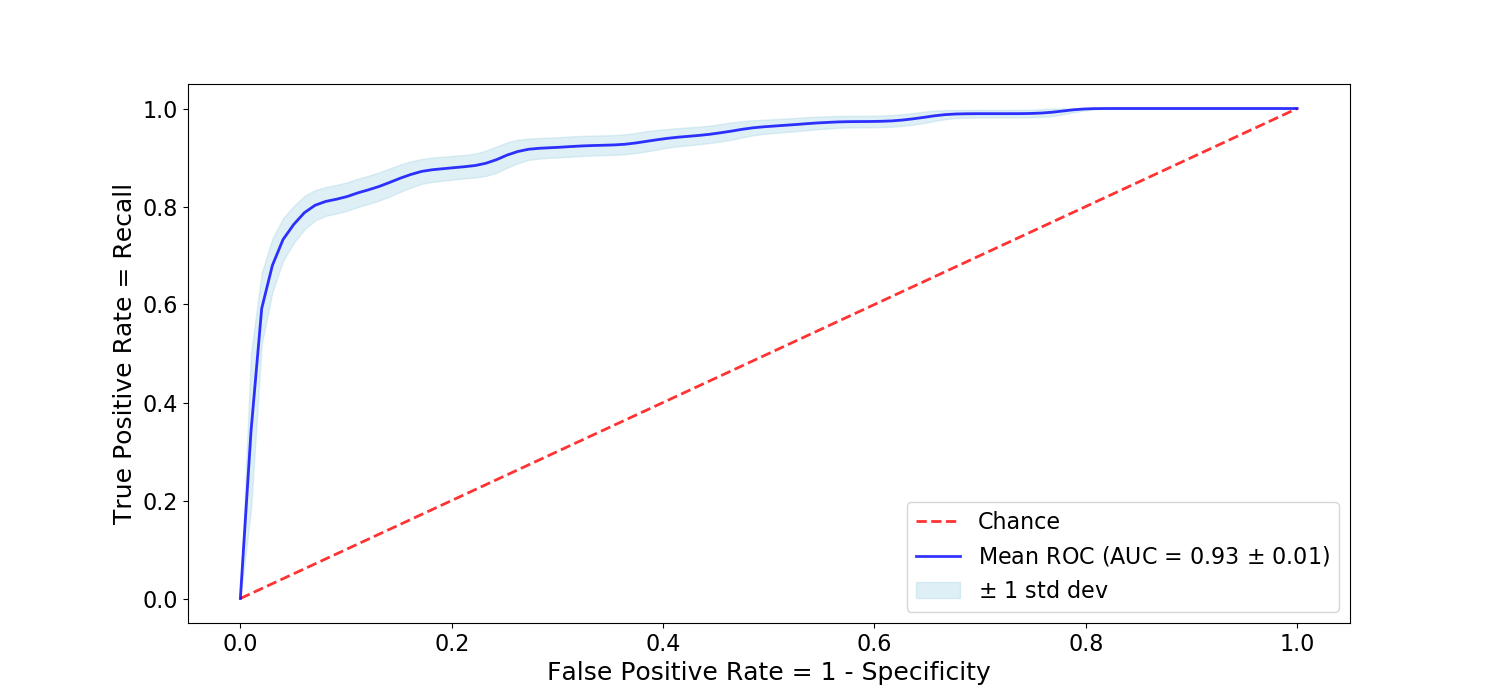}
	\caption{\protect\textbf{Vertebra-level fracture detection ROC curve} for the 5-fold cross-validation experiment, again bootstrapping (n=1000) our predictions to generate the mean and standard deviation of ROC curves (AUC=$0.93\pm0.01$).}
	\label{fig:roc_v}
\end{figure*}

\begin{figure}[!htb]
	\begin{subfigure}[b]{0.32\textwidth}
		\includegraphics[width=\textwidth]{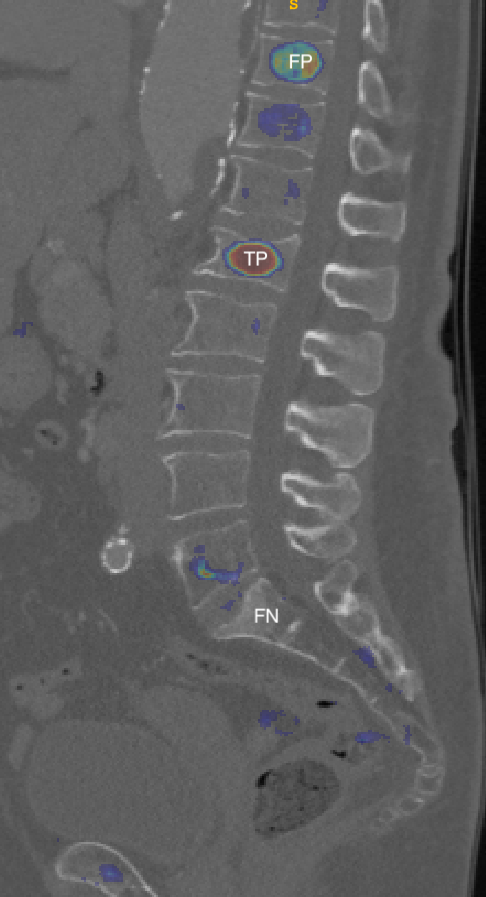}\hfill
		\caption{Validation case 3 with two errors at vertebra-level}
	\end{subfigure} \hfill
	\begin{subfigure}[b]{0.32\textwidth}
		\includegraphics[width=\textwidth]{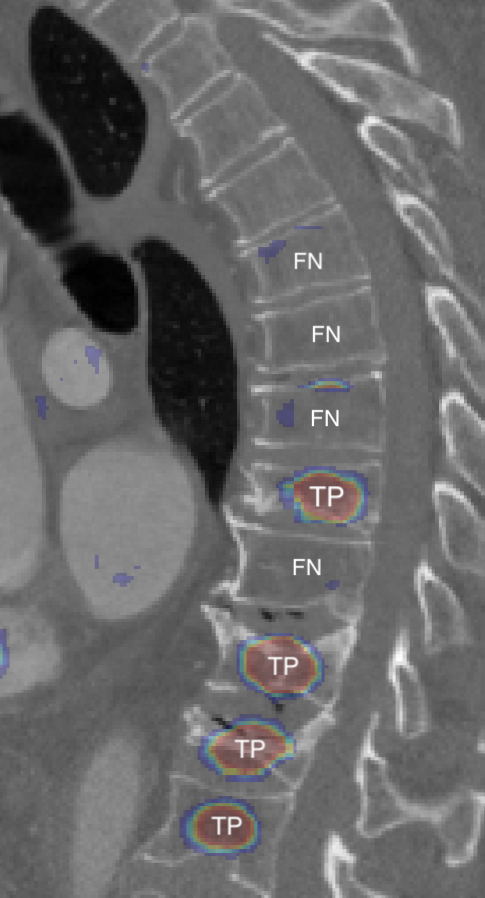}\hfill
		\caption{Validation case 4 with four FNs at vertebra-level}
	\end{subfigure} \hfill
	\begin{subfigure}[b]{0.32\textwidth}
		\includegraphics[width=\textwidth]{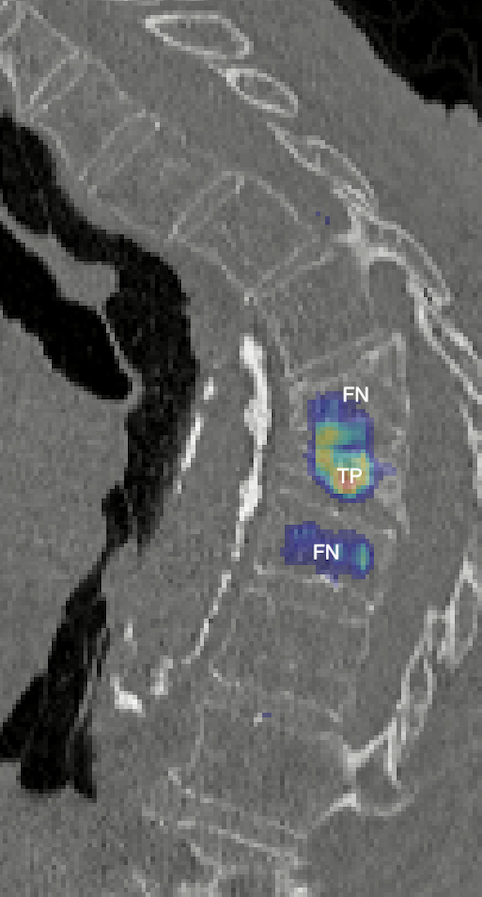}\hfill
		\caption{Validation case 5 with two FNs at vertebra-level}
	\end{subfigure}
	\caption{\protect\textbf{Fracture detection incorrect vertebra-level predictions} on three validation images: for each case, one mid-sagittal slice of the pre-processed 3D input image is overlayed with the output fracture class probability label map (blue = low probability, red = high probability), TPs and error type are annotated manually. Probability values \protect\textless0.05 have been removed for visualization purposes. Observe the differences in image quality and field of view present in our training database. Best viewed in color. (TP= True Positive, FP = False Positive, FN = False Negative)}
	\label{fig:pipeline2}
\end{figure}

Figure \ref{fig:roc_v} shows the \textbf{vertebra-level} fracture detection ROC curve for the five-fold cross-validation experiment (AUC of $0.93\pm0.01$). We added Gaussian noise to the ground truth vertebra coordinates (standard deviation of 3mm along each axis) to simulate using automatically detected centroid coordinates (see discussion in section \ref{sec:vert}). We are aware of one vertebral fracture detection work (using a 2.5D method~\cite{yao2012quantitative}) reporting a sensitivity of 95.7\% with a False Positive Rate of 0.29 per patient~\cite{burns2017vertebral}. Burns et al. designed their test set carefully by excluding cases with more than two contiguous vertebral fractures (in contrast, in our database 18\% of the cases contain more than two contiguous fractures and $>$70\% of vertebral fractures have at least one neighboring fractured vertebra, see Figure \ref{fig:pipeline2}(b) and (c)). We have not tested this 2.5D method on our test set because of the lack of an open source implementation.

%
%

\begin{figure}[!htb]
	\centering
	\begin{subfigure}[b]{0.4\textwidth}
		\includegraphics[width=\textwidth]{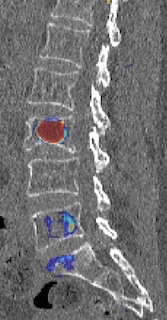}\hfill
		\caption{Validation case 1}
		\label{fig:a}
	\end{subfigure} \hfill
	\begin{subfigure}[b]{0.4\textwidth}
		\includegraphics[width=\textwidth]{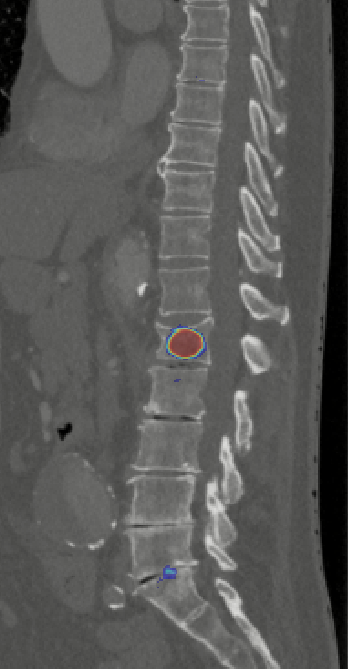}\hfill
		\caption{Validation case 2}
		\label{fig:a}
	\end{subfigure} \hfill
	\caption{\protect\textbf{Fracture detection correct vertebra-level predictions} on two validation images: for each case, one mid-sagittal slice of the pre-processed 3D input image is overlayed with the output fracture class probability label map (blue = low probability, red = high probability). Probability values \textless0.05 have been removed for visualization purposes. Best viewed in color.}
\label{fig:pipeline1}
\end{figure}

The vertebra-level fracture detection results are illustrated in Figure \ref{fig:pipeline1} (two validation cases with only correct vertebra-level predictions) and Figure \ref{fig:pipeline2} (three validation cases with False Positive (FP) and False Negative (FN) errors at vertebra-level). 
We observed that our vertebra-level errors occur predominantly on mild cases (either misses on ground truth mild fractures or false alarms on normal vertebrae) and can be clustered into the following categories: errors at edge vertebrae (vertebra and/or its neighbors are not completely visible, see Figure \ref{fig:pipeline2}(a)), errors in series of fractured vertebrae (known to be difficult to read as the reference vertebra \textit{dissappears}, see Figure \ref{fig:pipeline2}(b)) and errors due to confusion with other vertebra pathologies (e.g. inferior vertebra in Figure \ref{fig:pipeline2}(b)). Supported by Figure \ref{fig:data} we hypothesize that our training database does not contain enough vertebra examples, explaining for instance the FN on mild S1 in Figure \ref{fig:pipeline2}(a) and the FN on moderate T7 and mild T9 in Figure \ref{fig:pipeline2}(c) (notice the spine curvature around L5 and T7-T8 which makes these fractures look different compared to other locations along the spine). We also observed that some ambiguous (mild) cases would benefit from consensus reading as reported previously~\cite{buckens2013intra} (e.g. inferior vertebra in Figure \ref{fig:pipeline2}(b)).

\section{Conclusion}
\label{sec:concl}
We present to the best of our knowledge the first vertebral fracture detection model learning 3D features in the spine while simultaneously localizing the detection results to allow for interpretation by radiologists. We discussed the importance of exploiting 3D information to automatically learn compact vertebral fracture detection features. The results of our 5-fold cross-validation experiment demonstrate that our 3D data-driven method produces AUC scores above 90\% for patient-level and vertebra-level fracture detection.

While our work demonstrates encouraging fracture detection results, this study has a few limitations which can be mainly attributed to our small database. First, we reported vertebra-level fracture detection results with noisy manual annotations that should be replaced by automatically detected centroids (using an automated localization method). Secondly, we did not yet report fracture grades because our initial experiments show that we have an insufficient number of training examples for many vertebrae and especially for the more ambiguous mild fractures. Thirdly, the amount of thoracic vertebrae and the variability in spine pathologies and image acquisition settings was limited due to the size of our training database. Lastly, we used cross-validation instead of independent training and test sets due to the limited number of patients in our training database.

\FloatBarrier
\subsubsection*{Acknowledgements}
The authors thank the patients, investigators and their teams who took part in this study. The first author is grateful for the comments and feedback provided by Kasper Claes and the discussions on model evaluation with Roberto D'Ambrosio. This study was funded by UCB Pharma and Amgen Inc.

%
%
\bibliographystyle{splncs04}
\bibliography{refs}

\begin{thebibliography}{10}
\providecommand{\url}[1]{\texttt{#1}}
\providecommand{\urlprefix}{URL }
\providecommand{\doi}[1]{https://doi.org/#1}

\bibitem{bar2017compression}
Bar, A., Wolf, L., Amitai, O.B., Toledano, E., Elnekave, E.: Compression
  fractures detection on ct. In: SPIE Medical Imaging. pp. 1013440--1013440.
  International Society for Optics and Photonics (2017)

\bibitem{Bromiley2016}
Bromiley, P.A., Kariki, E.P., Adams, J.E., Cootes, T.F.: Fully automatic
  localisation of vertebrae in ct images using random forest regression voting.
  Lecture Notes in Computer Science  \textbf{10182},  51--63 (2016)

\bibitem{buckens2013intra}
Buckens, C.F., de~Jong, P.A., Mol, C., Bakker, E., Stallman, H.P., Mali, W.P.,
  van~der Graaf, Y., Verkooijen, H.M.: Intra and interobserver reliability and
  agreement of semiquantitative vertebral fracture assessment on chest computed
  tomography. PloS one  \textbf{8}(8),  e71204 (2013)

\bibitem{burns2017vertebral}
Burns, J.E., Yao, J., Summers, R.M.: Vertebral body compression fractures and
  bone density: Automated detection and classification on ct images. Radiology
  p. 162100 (2017)

\bibitem{genant1993vertebral}
Genant, H.K., Wu, C.Y., van Kuijk, C., Nevitt, M.C.: Vertebral fracture
  assessment using a semiquantitative technique. Journal of bone and mineral
  research  \textbf{8}(9),  1137--1148 (1993)

\bibitem{glocker2013vertebrae}
Glocker, B., Zikic, D., Konukoglu, E., Haynor, D.R., Criminisi, A.: Vertebrae
  localization in pathological spine ct via dense classification from sparse
  annotations. In: International Conference on Medical Image Computing and
  Computer-Assisted Intervention. pp. 262--270. Springer (2013)

\bibitem{Kamnitsas201761}
Kamnitsas, K., Ledig, C., Newcombe, V.F., Simpson, J.P., Kane, A.D., Menon,
  D.K., Rueckert, D., Glocker, B.: Efficient multi-scale 3d {CNN} with fully
  connected {CRF} for accurate brain lesion segmentation. Medical Image
  Analysis  \textbf{36},  61 -- 78 (2017)

\bibitem{koenig2006mevislab}
Koenig, M., Spindler, W., Rexilius, J., Jomier, J., Link, F., Peitgen, H.O.:
  Embedding vtk and itk into a visual programming and rapid prototyping
  platform. In: Medical Imaging 2006: Visualization, Image-Guided Procedures,
  and Display. vol.~6141, p. 61412O. International Society for Optics and
  Photonics (2006)

\bibitem{mader2017detection}
Mader, A.O., Lorenz, C., Bergtholdt, M., von Berg, J., Schramm, H.,
  Modersitzki, J., Meyer, C.: Detection and localization of landmarks in the
  lower extremities using an automatically learned conditional random field.
  In: Graphs in Biomedical Image Analysis, Computational Anatomy and Imaging
  Genetics, pp. 64--75. Springer (2017)

\bibitem{provost1997analysis}
Provost, F.J., Fawcett, T., et~al.: Analysis and visualization of classifier
  performance: Comparison under imprecise class and cost distributions. In:
  KDD. vol.~97, pp. 43--48 (1997)

\bibitem{tomita2018deep}
Tomita, N., Cheung, Y.Y., Hassanpour, S.: Deep neural networks for automatic
  detection of osteoporotic vertebral fractures on ct scans. Computers in
  biology and medicine  \textbf{98},  8--15 (2018)

\bibitem{valentinitsch2019opportunistic}
Valentinitsch, A., Trebeschi, S., Kaesmacher, J., Lorenz, C., L{\"o}ffler, M.,
  Zimmer, C., Baum, T., Kirschke, J.: Opportunistic osteoporosis screening in
  multi-detector ct images via local classification of textures. Osteoporosis
  International pp. 1--11 (2019)

\bibitem{waterloo2012prevalence}
Waterloo, S., Ahmed, L.A., Center, J.R., Eisman, J.A., Morseth, B., Nguyen,
  N.D., Nguyen, T., Sogaard, A.J., Emaus, N.: Prevalence of vertebral fractures
  in women and men in the population-based troms{\o} study. BMC musculoskeletal
  disorders  \textbf{13}(1), ~3 (2012)

\bibitem{yao2012quantitative}
Yao, J., Burns, J.E., Wiese, T., Summers, R.M.: Quantitative vertebral
  compression fracture evaluation using a height compass. In: SPIE Medical
  Imaging. pp. 83151X--83151X. International Society for Optics and Photonics
  (2012)

\end{thebibliography}
%
%
%
%
%

\end{document}